\documentclass[aip,pof,reprint,onecolumn,pdftex]{revtex4-1}
\usepackage{graphicx}%
\usepackage{amsmath}
\usepackage{amssymb}
\usepackage{eucal}
\usepackage{textcomp}
\usepackage{url}
\usepackage{color}

\begin{document}

\title{The distribution of ``time of flight'' in three dimensional stationary chaotic advection}
\author{Florence Raynal and Philippe Carri\`ere
} 
\affiliation{
LMFA, UMR CNRS -- Universit\'e de Lyon,\\
\'Ecole Centrale de Lyon-- Universit\'e Lyon 1 -- INSA Lyon,\\
\'Ecole Centrale de Lyon, 36 avenue Guy de Collongue, \\
69134 \'Ecully c\'edex, France.
}

\DeclareGraphicsExtensions{.png,.jpg,.pdf}

\begin{abstract}
The distributions of ``time of flight'' (time spent by a single fluid particle between two crossings of the Poincar\'e section) are investigated for five different 3D stationary chaotic mixers. 
Above all, we study the large tails of those distributions, and show that mainly two types of behaviors are encountered. 
In the case of slipping walls, as expected, we obtain an exponential decay, which, however, does not scale with the Lyapunov exponent. 
Using a simple model, we suggest that this decay is related to the negative eigenvalues of the fixed points of the flow. 
When no-slip walls are considered, as predicted by the model, the behavior is radically different, with a very large tail following a power law with an exponent close to $-3$. 
\end{abstract}

\maketitle


\section{Introduction}

Mixing in fluids comes with two mechanisms: \textit{stirring}, which consists in moving the fluid particles as efficiently as possible so as to create high gradients of concentration that are smoothed by \textit{molecular diffusion} thereafter.
Although mixing generally implies turbulent flows, it is now very well-known that chaotic advection enables  efficient stirring even when the flow is laminar \cite{bib:aref84,bib:ottino89}. 

Many recent studies, mostly in 2D flows, have shown how mixing in chaotic advection (and more especially the variance decay of a diffusive tracer) is controlled by regions of low stretching rate: 
in the presence of walls \cite{bib:chertkovlebedev03,bib:salmanhaynes07,bib:gouillartetal07,bib:gouillartetal08}, an algebraic decay of the variance, rather than the rapidly predominant exponential decay expected from early numerical simulations is observed \cite{bib:pierrehumbert94,bib:toussaintetal95,bib:antonsenetal96} (associated with the notion of
\textit{strange eigenmode} introduced by Ray Pierrehumbert \cite{bib:pierrehumbert94}, see also Giona \textit{et al.} \cite{bib:gionaetal04a,bib:gionaetal04b}).
More recently, in a 3D implementation of the randomized sine map \cite{bib:pierrehumbert94,bib:pierrehumbert00} (therefore without walls), Ngan and Vanneste \cite{bib:nganVanneste2011} suggest that the exponential variance decay is controlled by a few small fluid blobs that remain unstretched for long times. 

When dealing with realistic geometries of chaotic three dimensional flows, solving the advection-diffusion equation at high P\'eclet number is out of reach.
Because purely Lagrangian measures are easier to obtain \cite{bib:carriere07}, it is natural to search for a characterization of the influence of those regions of poor stretching in the usual tools of dynamical systems. 
The first idea which comes to mind is to consider \textit{Poincar\'e sections}: 
since the velocity field vanishes at fixed walls,
the density of points is lower in the vicinity of the walls than in the bulk;
but it is also lower in  regions  where the velocity component perpendicular to the Poincar\'e section vanishes.
The second simplest idea is to consider the \textit{Lyapunov exponents}, another classical tool of dynamical systems theory; 
but, as we will see in the paper, they fail to detect the presence of walls. 
Jones \& Young considered the \textit{axial dispersion} of a perfect or diffusive tracer along a twisted pipe \cite{bib:jonesyoung94}. 
For instance, they showed that in the case of the perfect tracer, the asymptotic ($t\rightarrow \infty$) shear dispersion $\sigma(t)$ varies like $t \ln t$ in the case of global chaos, whereas it varies like $t^2$ in other cases (not global chaos, regular regime or straight pipes); with a simple argument, they related this logarithmic behavior to the presence of walls. 
Then this measure of chaos is interesting since it can ``feel'' the presence of walls, while the other tools cited previously cannot. 
However, it has a major drawback when realistic geometries are under study: indeed, in order to detect the logarithmic behavior, they averaged 10 runs over long times, each run consisting of ensembles of $10^5$ particles. 
They used an analytical solution of their flow, which made the calculation ``affordable''. 
Otherwise, the computational cost would be too high for this parameter to be used systematically. 
\textit{Lobe dynamics} \cite{bib:romkedaretal90,bib:raynalgence95,bib:raynalwiggins2006} is a geometrical approach that gives interesting insights on mass exchange between different regions of the chaotic flow, but is quite restricted to 2D flows.
More recently, the \textit{linked twist map formalism} \cite{bib:sturman2006,bib:sturman2008}, available in 2 and 3D flows, has been proved to be a useful theoretical tool for design principles of efficient mixers available in many technological applications, and was extended for an idealized model of a class of fluid mixing devices of 2D flows to show how scalar decay is related to the presence of boundaries \cite{bib:SturmanSpringham2013}. 
Finally, the purely probabilistic \textit{transfer operator approach}, available in 2D and 3D flows, determines almost-invariant regions that minimally mix with their surroundings, and, unlike lobe dynamics, is able to detect regions with very small mass leakage \cite{bib:froylandpadberg2009}; 
the connection with \textit{topological chaos}  was done by Stremler \textit{et al.} \cite{bib:stremleretal2011}. 

In the present work, we propose to follow a simpler idea, that is to consider the \textit{time of flight}, lapse of time spent by a fluid particle between two consecutive crossings of Poincar\'e sections.
Since a fluid particle has a very slow motion when it is located in a region of low stretching, it spends more time between two crossings of the Poincar\'e section than it would otherwise, resulting in very long times of flight. 
As a particle wanders almost everywhere in the chaotic region, the histogram of times of flight may be considered as a global rather than local distribution; therefore an expected salient feature is that a satisfactory convergence (especially for the tail of the histogram) is obtained with only a reasonable amount of Poincar\'e section points (of order 10000, say), much less than for the axial dispersion discussed before. 
In practice, the histogram of times of flight may be smoothed by considering different initial points in the chaotic region, so as to obtain a reasonable tail for the statistics;
note however that the statistics (Lyapunov exponents, Poincar\'e sections) of each unique trajectory have to be sufficiently converged so that they do not depend on the choice of the initial point.

Using the time of flight is all the more interesting as it is already calculated in preparing the Poincar\'e section: 
once a chaotic mixer is under numerical study, it is expected at least to obtain a Poincar\'e section and see if chaos is global, so as to decide whether the mixer is efficient or not; 
the knowledge of the time of flight only requires to store the times at which the Poincar\'e section is crossed, or directly the difference between two consecutive crossing times.


\section{Time of flight and residence time distributions}

The distribution of time of flight has some reminiscences with the distribution of \textit{residence time} first introduced by Danckwerts \cite{bib:danckwerts53, bib:danckwerts58a},
a very usual tool in chemical engineering sciences; 
however, as we explain thereafter, they are definitively different.  

\subsection{Time of flight}
\label{sect:timeofflightdefinition}
As defined previously, the time of flight is the lapse of time between two crossings of Poincar\'e sections when following \textit{a single} fluid particle; 
it is directly connected to dynamical systems theory, since it is linked to the very definition of the Poincar\'e section. 
Let $P$ denote the Poincar\'e map: starting from an initial point located at $\mathbf{x}_0$ in the Poincar\'e section, 
the \textit{ordered} set of points is obtained as
\begin{equation}
	\mathbf{x}_n = P(\mathbf{x_{n-1}}),\ n \geq 0,
\label{eq:xn}
\end{equation}
where $n$ denotes the ordinal number of the Poincar\'e section when following the given trajectory (orbit), with an associated \textit{ordered} set of times of first return, $\tau_P$, in a time-continuous dynamical system (see Eckmann and Ruelle, section II.H \cite{bib:eckmannruelle85}):
\begin{equation}
	t_{\mathrm f}(n) = \tau_P(\mathbf{x_{n-1}}),\ n > 0\, .
\end{equation}
$t_{\mathrm f}$ is what we hereafter name ``time of flight",  while $\overline{t}_{\mathrm{f}}$ denotes the time of flight averaged over $n$. 

Note that we refer to a section in space. 
This may be contrasted with the time-periodic, 2D case, in which Poincar\'e sections based on the time-period are often used. 
Time of flight is intended for steady, 3D flows and, unlike the residence time (defined below), is a purely Lagrangian measure. 

Let us calculate the time of flight in an elementary flow like a cylindrical Poiseuille laminar flow: 
we suppose that two successive Poincar\'e sections are separated by a length $L$.
Because the flow is parabolic, a fluid particle will travel forever on the same straight streamline, at a velocity 
\begin{equation}
{\bf v}=v_x(r) {\bf e_x},\hbox{ with } v_x(r)=U_{\mathrm{max}}\,(1-(r/R)^2)
\label{vel_pipe}
\end{equation}
where $r$ is the radius at which the fluid particle is initially located, $U_{\mathrm{max}}=2U_{\mathrm{mean}}$ is the (maximum) velocity at the center of the pipe, and $U_{\mathrm{mean}}$ the mean velocity in a transverse section. 
Then the lapse of time between two crossings of Poincar\'e sections is always identical, equal to $L/v_x(r)$, and the corresponding time of flight distribution is a Dirac function at $t=L/v_x(r)$, only depending on the initial location of the given fluid particle.

Finally, note that the notion of time of flight is close to that of  waiting time \cite{bib:artuso_etal2008}, used in other branches of dynamical system community: 
the waiting time distribution $\psi_{\cal D}(t)$ over a domain ${\cal D}$ is the probability that a given particle entering ${\cal D}$ remains inside for a duration $t$ (waiting time); 
like the time of flight, it is a Lagrangian quantity, computed by running a \textit{single} long trajectory and recording waiting times. 
We will use the waiting time later in the paper.

\subsection{Residence time}
As defined by Danckwerts in his seminal paper of 1953 \cite{bib:danckwerts53}:\\
``Suppose some property of the inflowing fluid undergoes a sudden change from one steady value to another: for instance,
let the color change from white to red. Call the fraction of red material in the outflow at time $[t]$ later be $F([t])$."
The residence time distribution (RTD) $E(t)$ is the derivative of $F(t)$, as defined in equation (3) of his paper.
Note that 
\begin{equation}
\int_0^\infty E(t)\, dt=1,\ \hbox{ and that} \int_0^\infty t\;E(t)\, dt=t_{\mathrm{mean}}={\cal{V}}/q,
\end{equation}
where $\cal{V}$ is the volume of the mixer and $q$ is the flow-rate. 
Moreover, the residence time is an \textit{Eulerian} quantity, involving all the fluid (not just a single fluid particle) for the entire mixer (and not for a single slice of it). 

Danckwerts calculates RTD for a slice of cylindrical Poiseuille flow of length $L$ (in a non-diffusive case);
for $t>L/U_{\mathrm{max}}$ (minimal time needed by the fluid to appear at the outlet), we have:
\begin{equation}
E(t)=\frac{L^2}{2 U_{\mathrm{mean}}^2t^3}\ .
\end{equation}
Note finally that 
\begin{equation}
\int_{t_\textrm{min}}^\infty E(t)\, dt=1,\quad \text{ and that} \int_{t_\textrm{min}}^\infty t\;E(t)\, dt=t_{\textrm{mean}}=L/U_{\mathrm{mean}}
\end{equation}

One could wonder how to evaluate it in practice in a numerical work: 
a rather simple idea would be to seed some particles uniformly in the inlet section of the mixer \cite{bib:khakharOttino87,bib:mezic_etal99}, as for a pulse of concentration. 
In his other paper cited above \cite{bib:danckwerts58a}, Danckwerts shows, using a result from Spalding \cite{bib:spalding58}, that computing the RTD as a response of a pulse at inlet is only valid for a \textit{diffusive tracer}. 
However, following numerically a diffusive particle near a wall is tricky, since the particle is likely to end in the walls\dots  

Residence time is sometimes used as a generic term for many different quantities; 
in order to avoid confusion, this term will not be used in what follows. 


\section{Mixing systems and numerical approach}
\label{mixing_systems_and_numerical_approach}

In the following, we restrict our study to Stokes flows, and consider flows where chaos is global (no apparent regular regions, \textit{i.e.} the ergodic region covers the whole fluid domain), which are the cases of practical interest for efficient mixing. 
In order to investigate the time of flight distribution, we consider five different chaotic mixers, described in details later. 
The first one is the slipping wall cavity flow, for which an analytical solution exists. 
For all the other ones (another confined model flow and three realistic open-flow mixers, including the well known Kenics$^{\textrm{\textregistered}}$ 
\cite{bib:armeniadesetal66}), the flow-field is solved numerically via finite element method (FEM hereafter). 

The determination of time of flight distribution requires long asymptotic evaluations, which, in such complicated geometries, is a hard task. 
For instance, the loss of particles (that may end in the walls due to intrinsically limited numerical accuracy) must be as small as possible: 
indeed, particles with very long asymptotic time of flight are those that spend a lot of time near the walls.
Moreover, for the three open-flows, our results must not depend on the boundary conditions imposed at the inlet and the outlet.
Thus we checked our numerical results on two configurations: 
first of all, we simulated the first flow (the slipping wall cavity flow) via FEM, and found a perfect match with the results obtained with the analytical solution. 
In order to have more comparisons, we also used the Kenics$^{\textrm{\textregistered}}$, for which accurate numerical solutions are available in the literature. 
The numerical method, together with the method used for computing the Lyapunov exponents, are detailed in  Ref. \cite{bib:carriere06}; comparisons with other results (pressure loss, particle loss, etc.) for the Kenics$^{\textrm{\textregistered}}$ can be found in appendix \ref{app:num_kenics}: 
our results agree reasonably well with the existing literature. 

In the following, we briefly detail the different configurations and the results obtained in terms of Poincar\'e sections and Lyapunov exponents. 
Note that, strictly speaking, ``Poincar\'e section'' is somewhat improperly used here, although the extension is classical: for the cavity flows considered here, points with both positive or negative normal velocities are taken into account. 
For the in-line mixers, intersections are considered at points at cross-sectional planes, spaced according to the basic element, rather than following spatial periodicity, which would have twice this spacing.
Except when stated differently, hereafter, Lyapunov exponent means ``asymptotic'' Lyapunov exponents, by contrast with the so-called ``finite-time'' Lyapunov exponent we also discuss in the following. 
We recall that, in a steady 3-D flow, there are three ordered Lyapunov exponents of a Poincar\'e section
$\lambda_1 \geq \lambda_2 \geq \lambda_3$ satisfying
\begin{equation}
    \lambda_1 + \lambda_2 + \lambda_3 = 0,
\end{equation}
owing to incompressibility and
\begin{equation}
   \lambda_2 = 0,
\label{eq:lambda_2=0}
\end{equation}
because the dynamical system corresponding to particle fluid trajectory is continuous in time. It is easily deduced that:
\begin{equation}
	\lambda_3 = - \lambda_1 = - \lambda,
\end{equation} 
so that only the positive Lyapunov exponent (which may be zero) is required.
The Lyapunov exponent of the map $\widehat{\lambda}$ is then given by
\begin{equation}
     \widehat{\lambda} = \frac{\lambda}{\overline{t}_{\mathrm{f}}}\, .
\end{equation}

\subsection{Slipping wall cavity flow}
\label{sbsct:slipping_wall_cavity_flow}
\begin{figure}
\includegraphics[clip=]{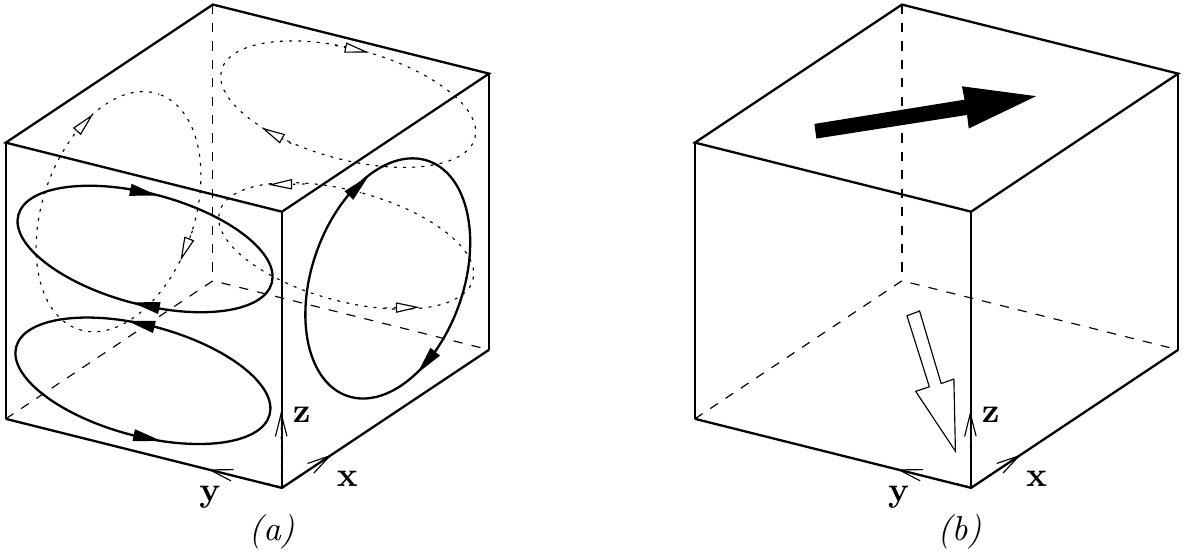}
    \caption{Sketch of the two confined flow-fields: \textit{(a)} the slipping wall cavity flow 
		\textit{(b)} the  no-slip wall cavity flow (with moving upper and lower walls); the second mimics the first more realistically.
	    }
		\label{fig:cube}
\end{figure}

\begin{figure}
\includegraphics[clip=]{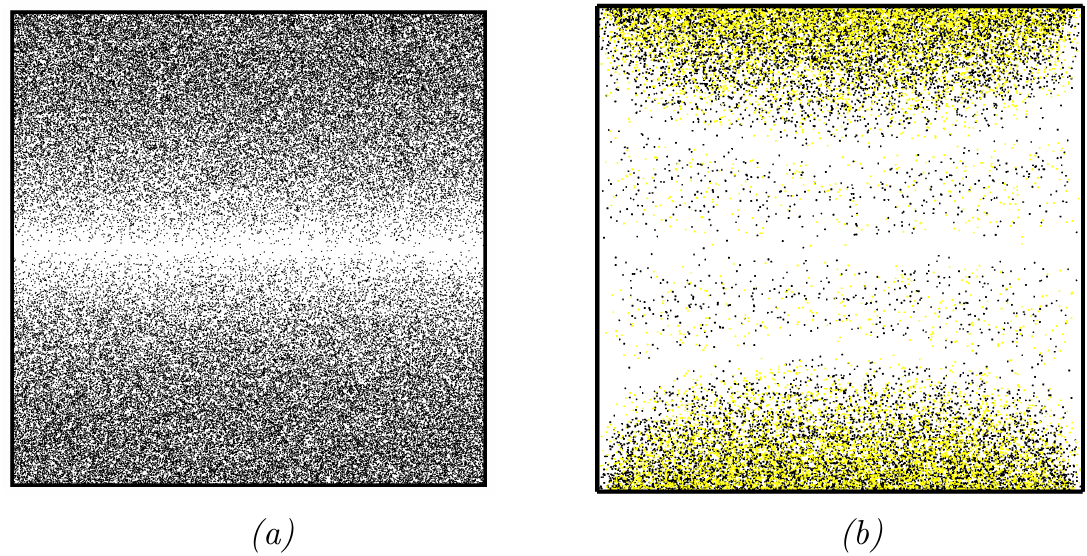}
    \caption{Poincar\'e sections (plane $(x=0.5)$) for: \textit{(a)} the slipping wall cavity flow in the case of global chaos $U_1=0.25$; 
		\textit{(b)} the no-slip walls cavity flow:
		in this case, two Poincar\'e sections are superimposed, represented by two different colors in the online version.
		\label{fig:poincaresectionTCRflows}
	    }
\end{figure}

The velocity-field is that of stationary 3-D flow in a cube with slipping boundaries, a case we have used in the past for numerical simulation of the advection-diffusion equation at high P\'eclet number \cite{bib:toussaintetal95,bib:toussaintcarriere99,bib:toussaintetal00}.
We recall that it is the sum of a steady main vortex, $\vec U_1$, of the Taylor kind whose axis is parallel to a side of the box, and of two counter-rotating steady plane vortices ($\vec U_2$) with equal amplitudes, see Figure \ref{fig:cube}$a$. 
The velocity field is:
\begin{eqnarray}
v_x&=& -U_1 \sin\pi x \cos\pi z \\
v_y&=& -2\,U_2 \sin\pi y \cos2\pi z \\
v_z&=& U_1 \cos\pi x \sin\pi z+ U_2 \cos\pi y \sin2\pi z 
\end{eqnarray}
where the constants $U_1$ and $U_2$ satisfy the normalization condition $U_1^2+5U_2^2/2=1$. 
We recall that this flow is globally chaotic for $U_1\le0.25$, and that values of $U_1$ such that $U_1\le0.15$ correspond to cases of global chaos with transadiabatic drift \cite{bib:bajermoffatt90}. 
The case $U_1= 0.25$ is the flow for which both chaos is global and the Lyapunov exponent is maximum ($\widehat{\lambda} = \ln 7.22$).
The corresponding Poincar\'e section (50,000 points here), calculated in a plane of constant $x$ passing through the center of the box, is reproduced in figure \ref{fig:poincaresectionTCRflows}.$a$.
The empty region near the middle plane corresponds to vanishing of the velocity component perpendicular to the Poincar\'e section. 

\subsection{No-slip walls cavity flow}
\label{sbsct:non-slipping_wall_cavity_flow}
In order to check the effect of a no-slip velocity field on the behavior of the time of flight 
distribution, we propose a second configuration, which mimics the preceding one, but in a more realistic manner:
the flow is induced by the stationary motion of the upper and lower walls ($z$ defining the vertical 
coordinate), co-moving in the $y$-direction and counter-moving in the $x$-direction \cite{bib:carriere06} (figure \ref{fig:cube}\textit{(b)}).
Somehow, it may be seen as a (stationary) three dimensional implementation of the time-periodic 2D cavity flow studied by
Leong and Ottino \cite{bib:leongottino89}.
As for the flow of the preceding section, Lagrangian properties depend on the relative amplitude of the
velocity components in the $x$- and $y$-direction; with the same ratio than herein, the chaotic region densely covers the whole domain. 
In this case, we expect an additional empty region in the Poincar\'e section at the vicinity of the fixed vertical walls. 
However, as can be noticed when looking at figure \ref{fig:poincaresectionTCRflows}.$b$, 
some more empty regions are visible:
two counter-recirculating vortices parallel to the $y$-axis are present rather than the single vortex of the slipping-walls case.
It may be inferred that the mixing efficiency of such a flow is lower than for the first one, with a Lyapunov number $\widehat{\lambda} = \ln 3.57$. 
Note finally that the number of section points we were able to calculate is lower than for the analytical flow: 
two sections are here superimposed, the first one with 14,734 points, the second one with 10,850, that clearly overlay each other. 

\subsection{Kenics$^{\textrm{\textregistered}}$ Mixer}

\begin{figure}
\includegraphics[clip=]{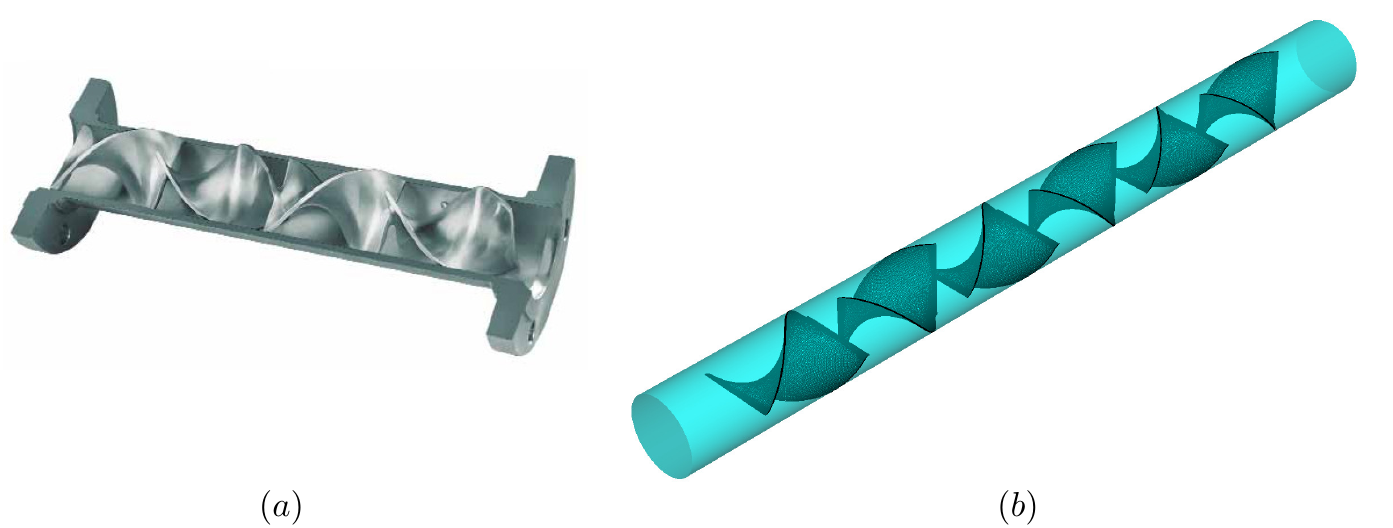}
    \caption{$(a)$ A Kenics$^{\textrm{\textregistered}}$ KM static mixer supplied by
		Chemineer, Inc.; $(b)$ Computational geometry.
		\label{fig:kenics_mixer_gmtry}
	    }
\end{figure}

The Kenics$^{\textrm{\textregistered}}$ mixer is probably the most famous and widely used static mixer. 
It is composed of a series of internal blades inside a circular pipe of
diameter $d$, each blade consisting of a short helix of length $L$ with a twist angle $\phi$.
The series is a succession of right- and left-handed blades, arranged alternately so that the leading edge of
a given blade is at right angle of the trailing edge of the preceding blade, thus with a spatial period of
length $2 L$.
A commercial model is shown in figure \ref{fig:kenics_mixer_gmtry}.$a$.
Hobbs and Muzzio performed simulations in this configuration using a commercial code for both flow simulation and particle tracking \cite{bib:hobbsmuzzio97} (see also Ref. \cite{bib:hobbsetal98}), while 
accurate numerical simulations for a large range of Reynolds number were performed by 
Byrde and Sawley \cite{bib:byrde97,bib:byrdesawley99,bib:byrdesawley99b}.
More references of experimental or numerical works are also available in the recent article by Kumar \textit{et al.}
\cite{bib:kumaretal08}.
In order to use part of the existing results as a check for our own calculations, we used 
the same parameters as O. Byrde \cite{bib:byrde97}, i.e. $L = 3d/2$ and $\phi = 180^{\mathrm{o}}$.

The geometry used for our FEM simulations is plotted in figure \ref{fig:kenics_mixer_gmtry}.$b$ 
Although a real mixer would involve about 12 or 16 successive blades, 
this 6-blades configuration is a good compromise between a realistic mixer and  
reasonable calculation time. 
\begin{figure}
    \begin{center}
      \includegraphics[clip=]{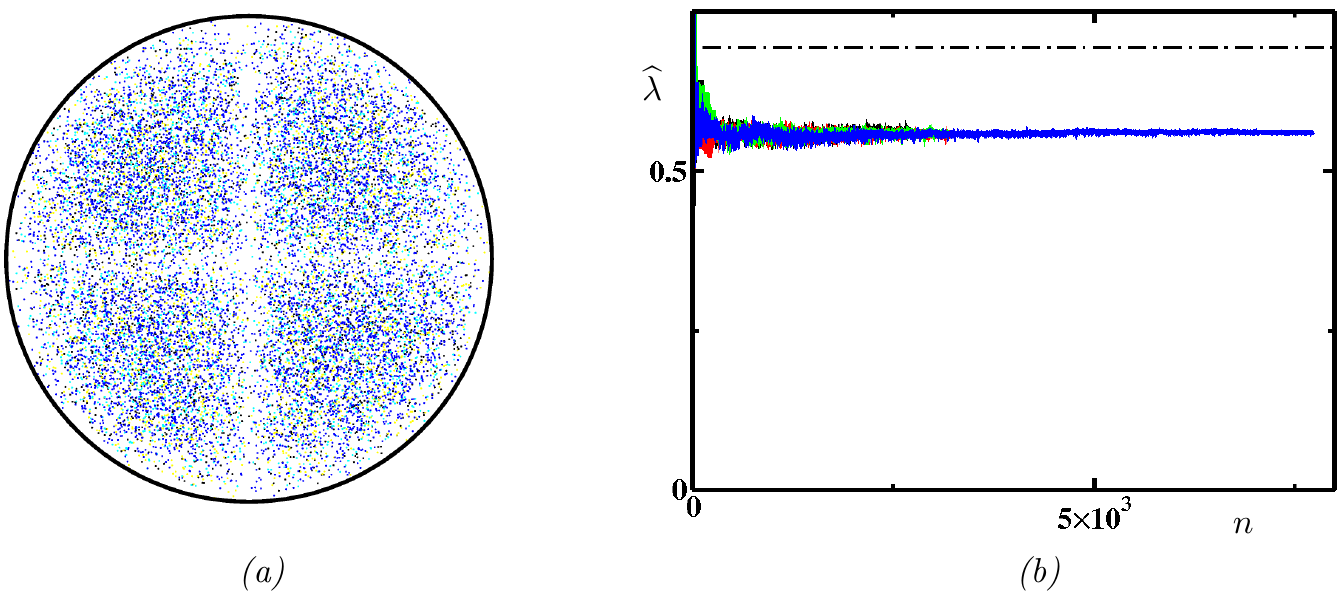}
    \caption{Kenics mixer: \textit{(a)} Poincar\'e sections (four sections are superimposed); 
		\textit{(b)} convergence of the positive Lyapunov exponent with the number of mixing elements $n$; the dot-dashed line is $\ln(2)$. 
		\label{fig:Kenics_Mixer_368951_r3-2_sct_lyap}
	    }
    \end{center}
\end{figure}
The Poincar\'e section (figure \ref{fig:Kenics_Mixer_368951_r3-2_sct_lyap}.$a$) shows a quite homogeneous global chaos away from the walls:
four Poincar\'e sections are superimposed on the plot, containing 2720, 3238, 3191 and 7737 points, respectively, corresponding to different initial locations, which clearly overlay each other. 

Note that the Lyapunov exponent converges towards $0.56 \approx \ln 1.75$ (figure \ref{fig:Kenics_Mixer_368951_r3-2_sct_lyap}.$b$), that is, a quite lower value than for the baker's map.
This is finally the point on which our result disagree the most with the existing literature.
Byrde and Sawley determined values slightly higher than $\ln 2$;
however, their calculations were performed in a context of non-negligible inertial effects (Reynolds number $25$ and $100$) which may enhance the resulting stretching: incidentally, the value they
predict for $R=100$ is largely higher than the one at $R=25$.
Also they dealt with \textit{finite time} Lyapunov exponents, that depend on the initial location of the particle, thus requiring some far from obvious averaging: 
At the opposite, the present \textit{asymptotic} Lyapunov exponents are naturally averaged over the domain and \textit{de facto} include the probability density function of each point in the Poincar\'e section. 
Note finally that our numerical simulations predict a rapid and clear convergence towards $\ln(2)$ for the two following static mixers (figure \ref{fig:baker_chen_meiners_sct}).
Thus the value of $\ln 1.75$ for a Stokes flow is indeed a measure of efficiency, and, in the case of creeping flows, mixing in this Kenics configuration is not as efficient as for the baker's map.

\subsection{Multi-level laminating mixer and "F" mixer}

\begin{figure}
    \begin{center}
      \includegraphics[clip=]{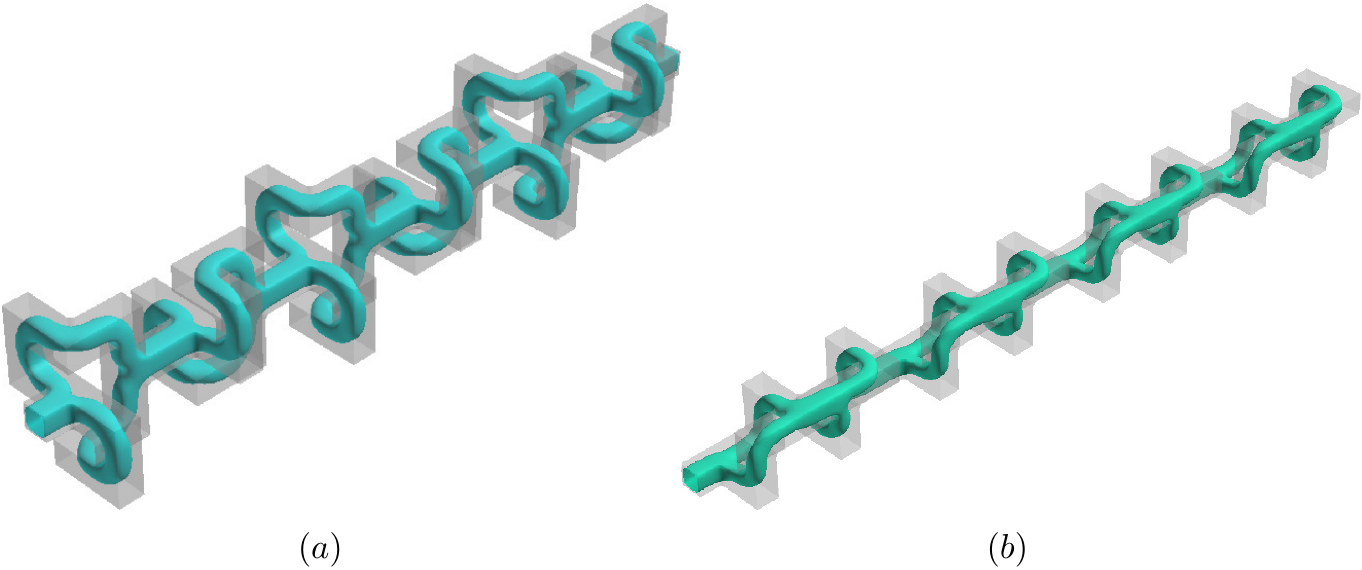}
    \caption{Geometry and iso-surface of velocity modulus plot for a Stokes flow inside: ($a$) the Multi-level laminating
		mixer (6 elements); ($b$) the "F" mixer of Chen and Meiners \cite{bib:chenmeiners04} (8 elements).
		\label{fig:baker_chen_meiners_gmtry}
	    }
    \end{center}
\end{figure}

In a previous paper \cite{bib:carriere07}, a three-dimensional flow configuration, which tries to mimic as close as possible the baker's map, was proposed and studied.
The corresponding geometry, here in the more realistic variant of an open flow composed of 6 basic ``mixing elements'',
is given in figure \ref{fig:baker_chen_meiners_gmtry}.$a$ together with a plot of an iso-surface of velocity modulus for
illustrating the separation--stacking process.
The design is close to the multi-level laminating mixer (MLLM) proposed by Gray \textit{et al.} \cite{bib:grayetal99}.
The successive elements are inverted so as to break the symmetry of the flow and
eliminate small residual islands in the Poincar\'e section. 
Such a mixer configuration is sometimes named ``baker's flow'' \cite{bib:lesteretal2012,bib:lesteretal2013}. 
Because the results, in the present context, are very similar, we present simultaneously the case of the ``F'' mixer of Chen and Meiners \cite{bib:chenetal09,bib:chenmeiners04},  whose geometry is given in figure \ref{fig:baker_chen_meiners_gmtry}.$b$. 
\begin{figure}
    \begin{center}
\includegraphics[clip=]{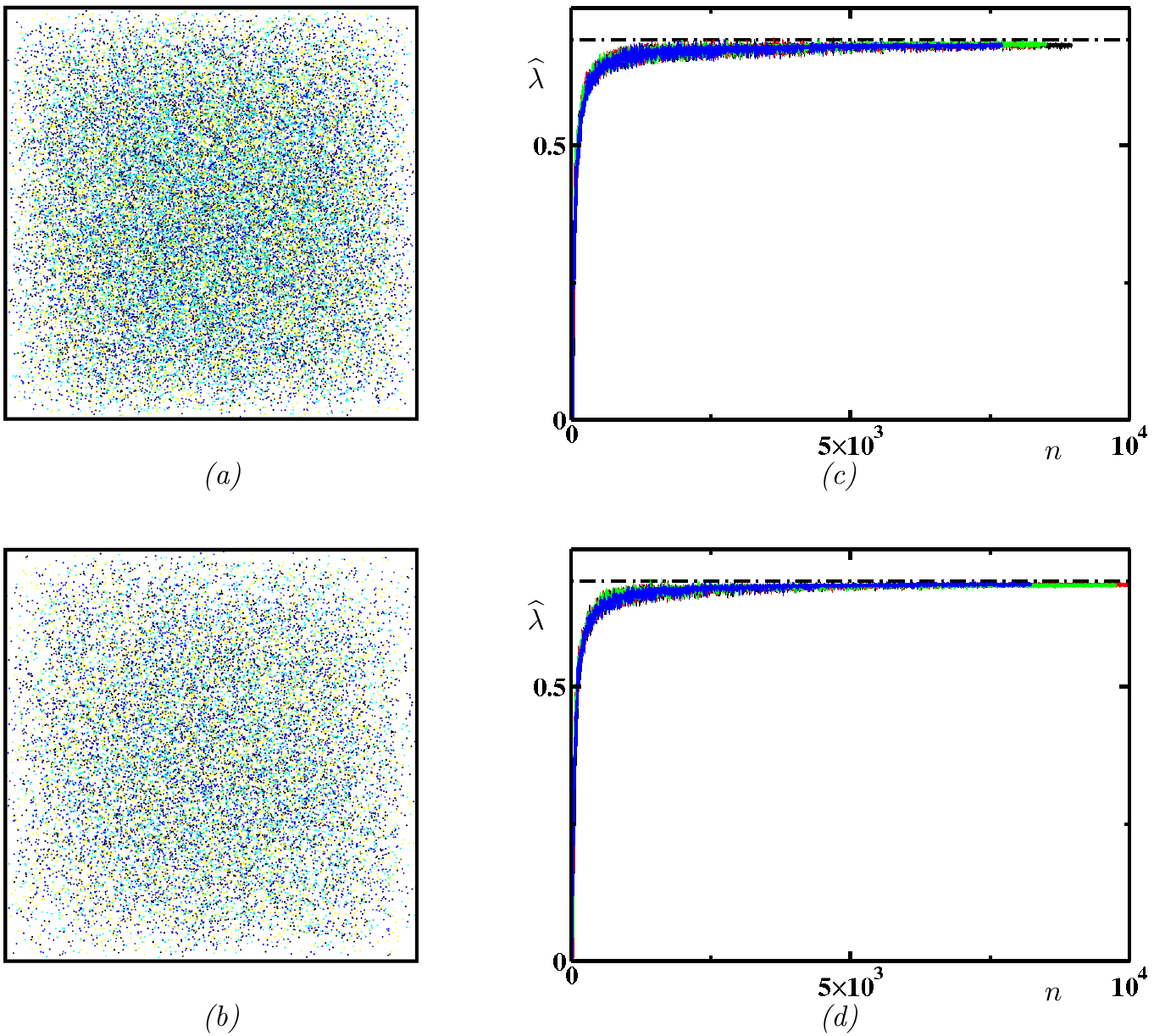}
    \caption{Left: Poincar\'e sections (four sections are superimposed) for ($a$) the Multi-level laminating mixer; 
		($b$) the ``F'' mixer.
		Right: convergence of the positive Lyapunov exponent with the number of mixing elements $n$ for
		($c$) the Multi-level laminating mixer; ($d$) the ``F'' mixer.
		\label{fig:baker_chen_meiners_sct}
	    }
    \end{center}
\end{figure}
It may appear somewhat surprising to retrieve that, as alluded before, the Lyapunov exponent is $\ln 2$ within the accuracy of the numerical method (figures \ref{fig:baker_chen_meiners_sct}.$c$ and \ref{fig:baker_chen_meiners_sct}.$d$). 
Thus, despite the walls, these two mixers succeed in approaching the mean behavior of the baker's map. 
Figure \ref{fig:baker_chen_meiners_sct} shows Poincar\'e sections for each mixer: 
four Poincar\'e sections having 8959, 8387, 8508 and 7716 points are superimposed in Fig. \ref{fig:baker_chen_meiners_sct}.$a$ and sections with 5000, 4989, 4877 and 4121 points in \ref{fig:baker_chen_meiners_sct}.$b$.

\section{Time of flight}

\begin{figure}
  \begin{center}
\includegraphics[clip=]{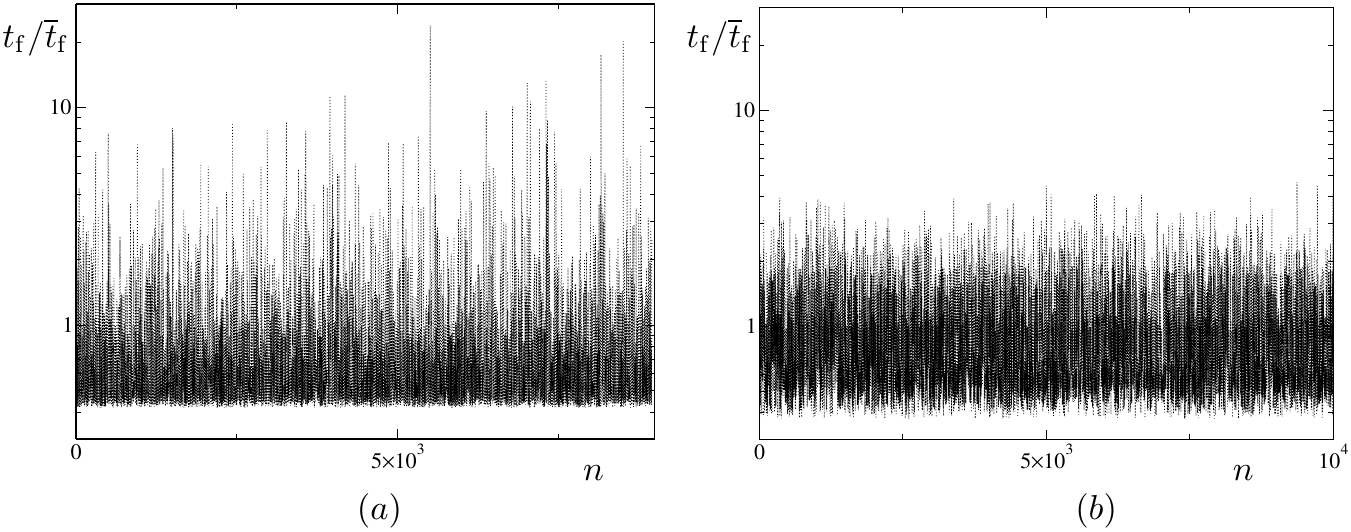}
  \caption
    {
      Non-dimensional time of flight, $t_{\mathrm{f}}/\overline{t}_{\mathrm{f}}$, versus the ordinal number of
      Poincar\'e section points, $n$ for a unique trajectory,
      corresponding to:  $(a)$ one of the Poincar\'e sections in figure
      \ref{fig:baker_chen_meiners_sct}.$a$ (multi-level laminating mixer);
      $(b)$ the trajectory in figure \ref{fig:poincaresectionTCRflows}.$a$ (slipping-walls cavity flow).	
      \label{fig:timeofflightvspsctpt} 
    }
  \end{center}
\end{figure}

\begin{figure}
  \begin{center}
\includegraphics[clip=]{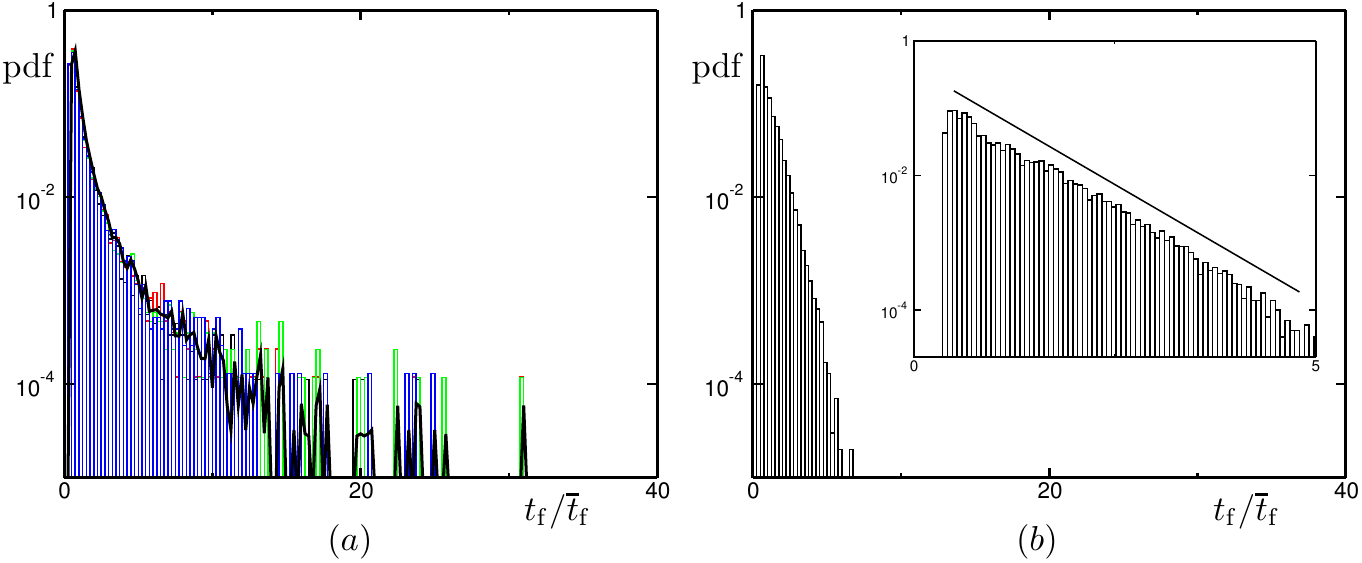}
   \caption
    {
	Distribution of reduced time of flight for: $(a)$ the multi-level laminating mixer; 
	$(b)$ the slipping-walls cavity flow.
	In the former case, four histograms corresponding to the four trajectories of
	figure \ref{fig:baker_chen_meiners_sct}.$a$ are superimposed; the
	thick black solid line represents the distribution averaged over
	these four histograms. 
        In the second case, the distribution is shown at another scale, so as to see clearly the exponential decay.
        \label{fig:ToF_distribution_log_lin}
    }
  \end{center}
\end{figure}

As defined earlier, time of flight $t_{\mathrm{f}}$ is the time spent by a single particle 
between two consecutive intersections with the Poincar\'e  plane.
We denote by $n$ the ordinal number of Poincar\'e section when following the given trajectory, and by $\overline{t}_{\mathrm{f}}$ the time of flight averaged over $n$. 
Figure \ref{fig:timeofflightvspsctpt}$a$ shows a typical plot of the behavior of the ``reduced time of flight'' $t_{\mathrm{f}}/\overline{t}_{\mathrm{f}}$ as a function of $n$ in a mixer with fixed no-slip boundaries (here the multi-level laminating mixer). 
As expected for a chaotic trajectory, it exhibits highly random behavior. 
Note the use of a logarithmic scale for the vertical axis, so as to allow for extreme events (large departures from the mean), corresponding to situations where the particle is trapped for a long time in the near vicinity of walls, before escaping to the core of the flow.
In this respect, it is clear that the statistics in cases $a$ and $b$ (respectively no-slip and slipping boundaries) are dissimilar.
As a consequence, the tails of the distributions of time of flight are expected to be quite different depending on the presence or not of no-slip walls:
in figure \ref{fig:ToF_distribution_log_lin}, we compare the probability density functions (pdf) of the reduced time of flight $t_{\mathrm{f}}/\overline{t}_{\mathrm{f}}$ for the two preceding cases; the two pictures are plotted with the same lin-log scale.
As one would expect in a globally chaotic flow, the distribution of time of flight indeed reveals an exponential decay with $t_{\mathrm{f}}/\overline{t}_{\mathrm{f}}$ in the slipping wall cavity flow, but the result is completely different concerning the no-slip walls flow. 
Thus we will consider those two cases separately thereafter. 

\subsection{No-slip boundaries}
\subsubsection{Theoretical model}
\begin{figure}
  \begin{center}
\includegraphics[clip=]{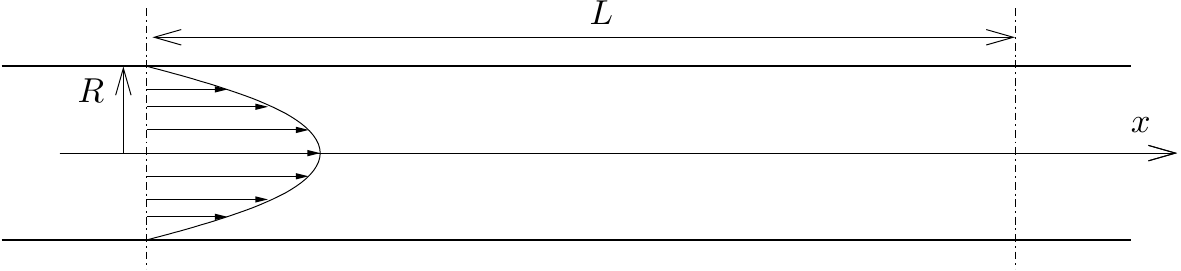}
    \caption
    {
    Poiseuille flow: the time of flight is calculated with Poincar\'e sections being separated by a length $L$.
        \label{fig:Poiseuille}
    }
  \end{center}
\end{figure}

In order to understand this non-exponential behavior in the presence of walls, we propose to mimic the trajectory of a single fluid particle in such chaotic flows as follows:
\begin{itemize} 
\item the flow in an element of the mixer is modeled by a non-chaotic flow, here possessing no-slip boundaries;  
\item the effect of global chaos on the trajectory of the fluid particle is modeled by \textit{random reinjection} at the entry to the next element \textit{with a non uniform probability distribution}, that takes into account the fact that the particle randomly samples the whole section, but less near the walls; 
\item in order to preserve mass conservation, \textit{the probability density function of the location of reinjection is taken proportional to the velocity rate}.
\end{itemize}
For instance, a mixing element of the Kenics$^{\textrm{\textregistered}}$ is replaced by a piece of cylindrical pipe, or the no-slip walls cavity flow is modeled by a piece of plane Couette flow; 
in each element of the model, the trajectory is thus a straight segment following a streamline of the flow, while the location of the particle changes at each new element. 
The shape of the distribution can be further explained as follows:
during a lapse of time $dt$, less particles of the flow cross the section near the walls than in the core where the velocity is maximum;  therefore the probability density for the single particle to cross the section at a given point must also follow this flux of particles. 
This last property was also used in a 3D-model of chaotic flow with sources and sinks in a Hele-Shaw cell, where the flow was calculated first in 2D, and the $z$-dependence was modeled by a parabolic reinjection rate from the source, with surprisingly good agreement between the model and 3D-calculations \cite{bib:raynaletal2013}. 
Using these model flows, it is possible to obtain an analytical expression for the distributions of time of flight. 
We present hereafter the calculations in a cylindrical pipe (Poiseuille flow, see figure \ref{fig:Poiseuille}), with velocity-field given by equation (\ref{vel_pipe}).
The calculation for the plane Couette flow is developed in Appendix \ref{app:analytic_timeofflight}. 

Let $G(t)$ be the probability density to have a time of flight of duration $t$ for an element of length $L$; 
$G(t) dt$ is therefore the probability to have a time of flight in between $t$ and $t+dt$. 
Given
\begin{equation}
t=L/v_x(r)\; ,
\label{timeflight_pipe}
\end{equation}
$t$ depends only on $r$, so this probability is equal to that of having a particle between $r$ and $r+dr$, with $r$ the radius related to time of flight $t$ by equation (\ref{timeflight_pipe}). 
Thus the probability of having a particle reinjected in between $r$ and $r+dr$ is such that 
\begin{equation}
G(t) dt \propto v_x(r) 2\pi r dr.
\end{equation}
From equations  \ref{vel_pipe} and \ref{timeflight_pipe}, we have:
\begin{equation}
1-(r/R)^2 =\frac{L}{U_{\mathrm{max}}\, t}
\end{equation}
that can be differentiated into
\begin{equation}
-2r/R^2 dr=-\frac{L}{U_{\mathrm{max}}\, t^2}dt.
\end{equation}
We finally obtain:
\begin{equation}
G(t)\propto t^{-3}.
\end{equation}
We also obtain a $t^{-3}$ tail in the case of plane Couette flow (see Appendix  \ref{app:analytic_timeofflight}), and plane Poiseuille flow (calculation not given here).

\subsubsection{Time of Flight histograms}
\begin{figure}
  \unitlength = 1.0cm
  \begin{center}
\includegraphics[clip=]{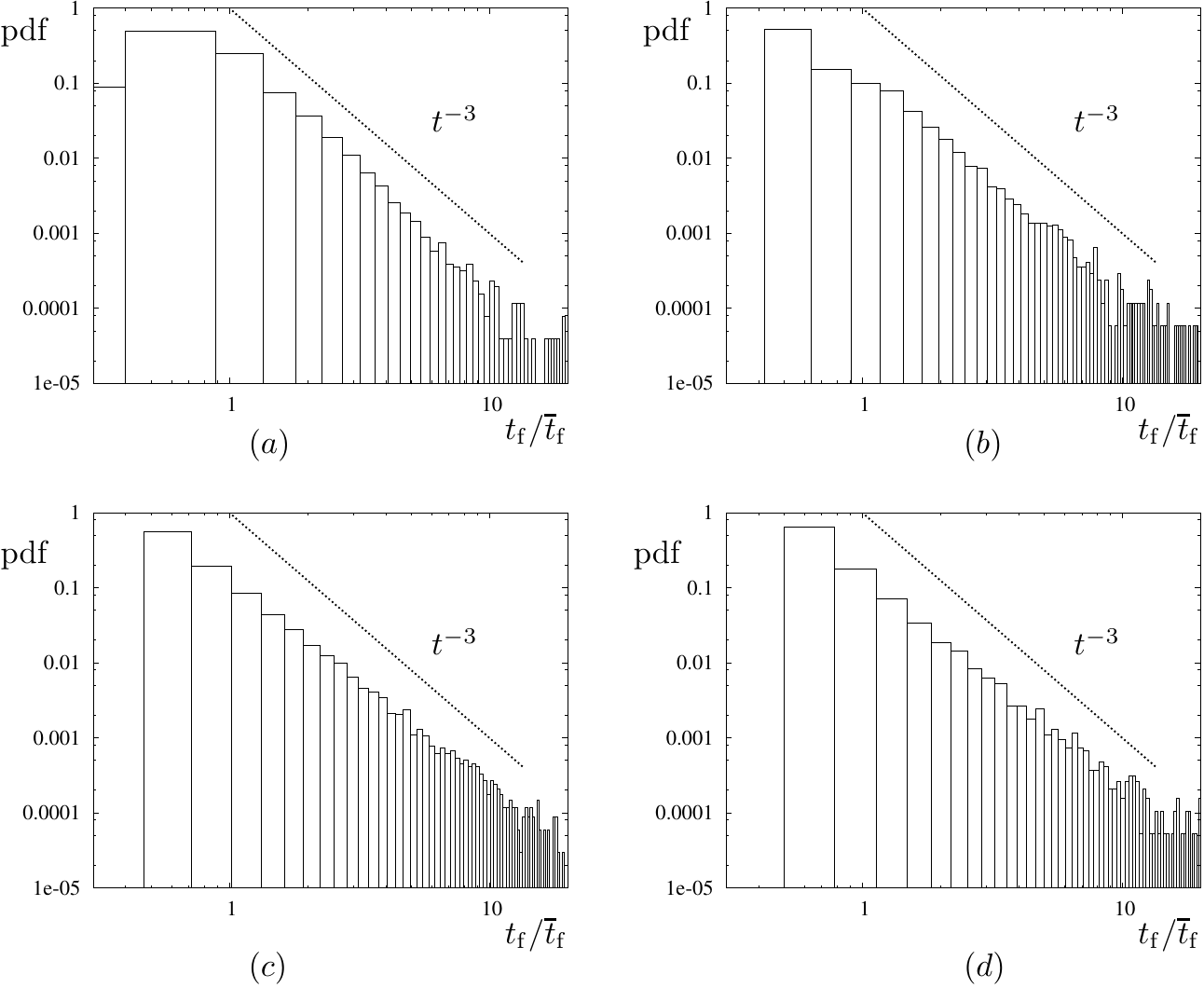}
    \caption
    {
	Distribution of time of flight for: $(a)$ the \textit{no-slip walls} cavity flow; 
	$(b)$ the Kenics$^{\textrm{\textregistered}}$ mixer; $(c)$ the multi-level laminating mixer; $(d)$ the "F" mixer.
	The histogram is averaged over two trajectories (used to plot figure \ref{fig:poincaresectionTCRflows}.$b$) in plot $(a)$, and those in plots $(b$--$d)$ are averaged over four trajectories (similarly those used to plot the corresponding Poincar\'e sections); 
        in all cases the smallest count is the inverse of the number of times of flights calculated, of the order of:
        $(a)$: $4\, 10^{-5}$; $(b)$: $6\, 10^{-5}$; $(c)$: $3\, 10^{-5}$; $(d)$: $5\, 10^{-5}$. 
	The dotted line stands for the $t^{-3}$ power law.
        \label{fig:ToF_distribution_log_log}
    }
  \end{center}
\end{figure}
In order to compare the predictions from our model with our numerical results, we plot in figure \ref{fig:ToF_distribution_log_log} the histograms of time of flights (calculated together with the Poincar\'e sections shown in figures \ref{fig:poincaresectionTCRflows}$b$, \ref{fig:Kenics_Mixer_368951_r3-2_sct_lyap}.$a$, and \ref{fig:baker_chen_meiners_sct}.$a$-$b$ respectively), in log-log scales, for all the mixers with no-slip boundaries described in section \ref{mixing_systems_and_numerical_approach} (namely, the no-slip walls cavity flow, the multi-level laminating mixer, the ``F'' mixer, and the Kenics$^{\textrm{\textregistered}}$ mixer):
even in the case of the no-slip walls cavity, which may be considered as different from the three more realistic static mixers, the histograms exhibit a power law with an exponent close to $-3$. 
Although the details of the flow may influence the short time statistics, and therefore, because of mass conservation, the amplitude of the tail, note that those histograms have a very similar shape, with close values of absolute amplitude of the algebraic tails. 
The fact that we recover the same type of behavior for the distribution of time of flight from numerical results and with our model favors the hypothesis that the shape of the distribution of time of flight is a signature of the presence (or not) of solid fixed walls inside the flow.


\subsection{The slipping-walls (TCR) flow}
As shown in figure \ref{fig:timeofflightvspsctpt}.$b$, the tail of the histogram is clearly exponential, as one would expect in a fully chaotic flow; however, it does not scale with the Lyapunov exponent. 
Indeed, the Lyapunov exponent can be seen as a ``mean stretching rate'', that takes into account regions of high or low stretching visited by the fluid particle, while the tails of histograms correspond to long time of flights, connected to trapping of the particle in regions of low stretching rates.
Thus the reason for this exponential decay is not completely entangled in the chaotic nature of the flow, but rather may be explained by the presence of hyperbolic fixed points:
it requires infinite time for a point located \textit{exactly} on the stable manifold of an hyperbolic fixed point to reach this fixed point; 
thus it may take arbitrary long time for a fluid particle very close to the stable manifold to reach the vicinity of the fixed point before escaping along the unstable manifold. 
Those ``trappings'' along a stable manifold, although scarce, may lead to those rare long time events for the time of flight. 
Simulations available as supplementary material \cite{bib:carriere02a,bib:carriere02b} support this hypothesis.

\subsubsection{Theoretical model}
\begin{figure}
  \begin{center}
\includegraphics[clip=]{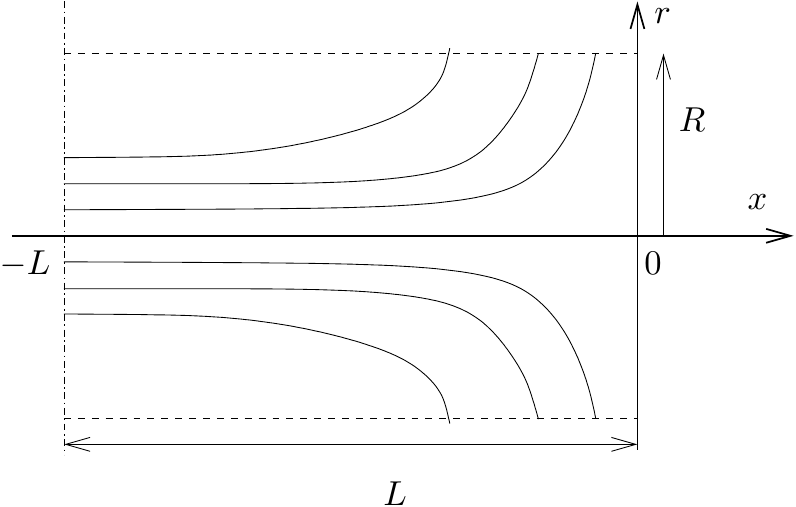}
   \caption
    {
    Flow in the vicinity of an hyperbolic point (located here at $(x=r=0)$). We define a domain ${\cal D}=(-L\le x\le0)\times(r\le R)$ around this fixed point; the waiting time in ${\cal D}$, denoted by $\tau$, is the time it takes to move from the plane $(x=-L)$ up to the cylinder $(r=R)$.
        \label{fig:Flow_fixed_point}
    }
  \end{center}
\end{figure}

If long times of flight are due to a trapping near a fixed point of the flow, then distributions of times of flight $t_{\mathrm f}$ have the same long time behavior as \textit{waiting times} $\tau$ (defined at the end of paragraph \ref{sect:timeofflightdefinition}) in a domain around this fixed point. 
Similarly to the case of flows with walls, we propose a model flow that evaluates the waiting time in the vicinity of a fixed point, constructed as follows:
\begin{itemize}
\item the flow around a fixed point is modeled by a non-chaotic flow in a domain ${\cal D}=(-L\le x\le0)\times(r\le R)$, here possessing a hyperbolic fixed point, like depicted in figure \ref{fig:Flow_fixed_point}; 
\item Chaos is modeled by a random reinjection in the plane $(x=-L)$, with reinjection probability distribution proportional to the local velocity rate.
\end{itemize}
Using this model, we can first calculate the waiting time of a given fluid particle, then obtain the corresponding distribution. 

\underline{Waiting time $\tau$ of a particle with given entrance location $r=r_0$:}
in the domain $\cal{D}$, the velocity field is well-described by the equations
\begin{eqnarray}
v_x&=& -\lambda x\label{eq:vx}\\
v_r&=&\lambda r/2\, .\label{eq:vr}
\end{eqnarray}
Consider an individual particle that enters the domain ${\cal D}$ at $t=0$ at point $(x_0,r_0)$, and leaves it at $t=\tau$ at point $(x(\tau),r(\tau))$. 
Using the boundary conditions $x_0=-L$ and $r(\tau)=R$, and equations (\ref{eq:vx}) and (\ref{eq:vr}), we obtain:
\begin{eqnarray}
x(\tau)&=& -L\, \exp(-\lambda\, \tau)\label{eq:x_t_f}\\
R&=& r_0\, \exp(\lambda \tau/2)\label{eq:r_t_f}
\end{eqnarray}
so that $\tau$ is well defined by the knowledge of $r_0$ using equation (\ref{eq:r_t_f}).

\underline{Distribution of waiting times in the domain ${\cal D}$:}
let $\psi_{{\cal D}}(\tau)$ be the probability density to have a waiting time $\tau$, and $\psi_{{\cal D}}(\tau) d\tau$ the probability to have a waiting time in between $\tau$ and $\tau+d\tau$. 
Thus $\psi_{{\cal D}}(\tau)$ verifies, for all particles entering the domain at $x_0=-L$ through the circlet between $r=r_0$ and $r=r_0+dr_0$: 
\begin{equation}
\psi_{{\cal D}}(\tau)\, d\tau\propto v_x\vert_{_{x=-L}}\, 2\pi r_0\, dr_0\, ;
\label{eq:E_enter}
\end{equation}
from equation (\ref{eq:vx}), and using $dr=(\lambda/2)\, r d\tau$ (from equation (\ref{eq:vr})), we obtain
\begin{equation}
\psi_{{\cal D}}(\tau) \propto  L\lambda^2\pi\, r_0^2,
\end{equation}
and finally, from equation (\ref{eq:r_t_f})
\begin{equation}
\psi_{{\cal D}}(\tau) \propto \pi R^2\, L\,\lambda^2 \exp(-\lambda \tau).
\label{eq:E_propto_exp}
\end{equation}

Note that the same analysis, carried on particles that \textit{leave} $\cal{D}$ at $r=R$, changes equation (\ref{eq:E_enter}) into
\begin{equation}
\psi_{{\cal D}}(\tau)\, d\tau\propto v_r\vert_{_{r=R}}\, 2\pi R\, dx\, ,
\label{eq:E_leave}
\end{equation}
which, using equations (\ref{eq:x_t_f}) and (\ref{eq:vx}), naturally leads to the same result as in equation (\ref{eq:E_propto_exp}). 

\underline{Times of flights:}
as explained before, we are mainly interested in the long-range decay of $t_{\mathrm f}$ (long times of flights), for which we can consider that $t_{\mathrm f}\sim \tau$.  
Thus the time of flight $t_{\mathrm f}$ should also have an exponential probability distribution, scaling with negative eigenvalue of fixed point:
\begin{equation}
 G(t)\propto \exp(-\lambda t)\, .
\end{equation} 

\subsubsection{Time of flight histograms}
In the model above we have shown that the time of flight distribution in the presence of a single hyperbolic point should decay exponentially, following the negative eigenvalue of this given fixed point. 
However, in the whole flow, there are \textit{many} different fixed points, associated with \textit{many} different negative eigenvalues. 
We could wonder therefore what the time of flight histograms will look like.

\begin{figure}
  \unitlength = 1.0cm
  \begin{center}
\includegraphics[clip=]{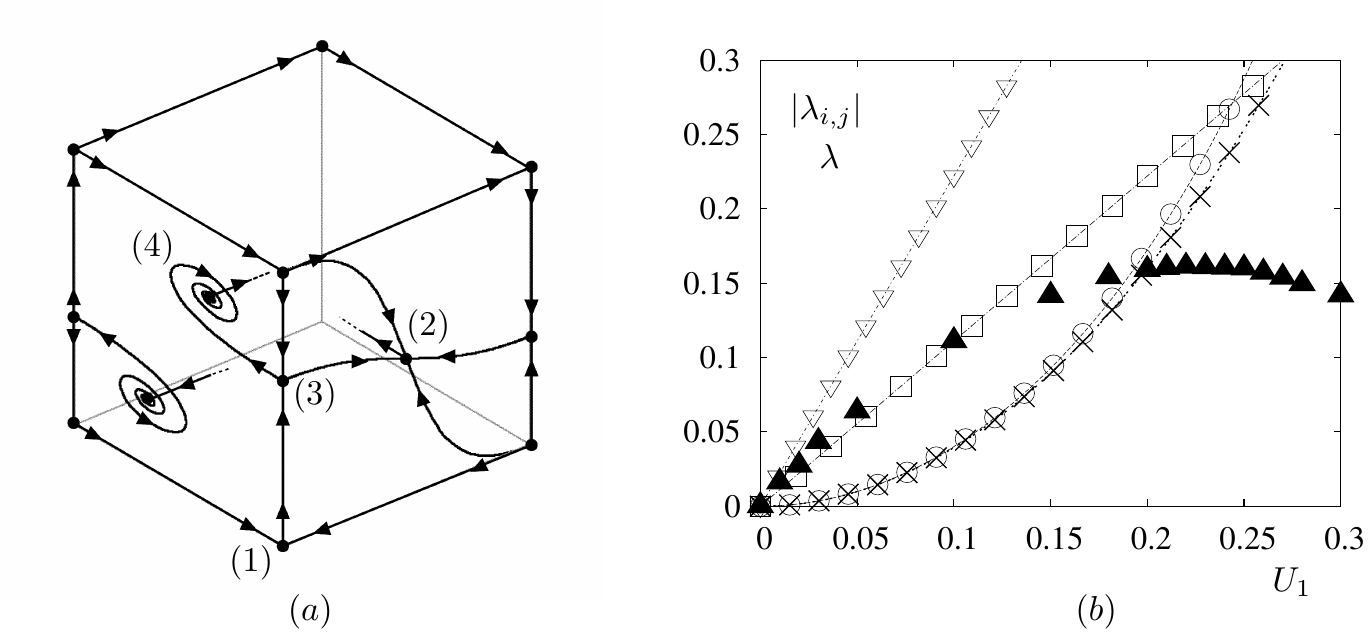}
    \caption{
	$(a)$: sketch of fixed points in the case of global chaos $U_1=0.25$.
	Fixed points are marked with a filled circle and stable/unstable manifolds
 are indicated with arrows; for``non-trivial'' manifolds, not located on the sides of the cubes, only a very small piece is drawn, ended by a dashed line. 
	Fixed points belonging to the rear sides are omitted for sake of clarity, and may be deduced from symmetry arguments. 
A fixed point denoted by $(i)$ ($i=1,2,3$ or $4$) on the figure has eigenvalues named $\lambda_{i,j}$ thereafter. 
	$(b)$: absolute values of negative eigenvalues of fixed points for $U_1\le0.3$ from appendix \ref{app:fixed_points_slip}, together with the Lyapunov exponents of the flow. For sake of clarity, only those nearer to zero are shown in the figure. ($\blacktriangle$): Lyapunov exponent $\lambda$; ($\times$): $\lambda_{3,1}$; ($\square$): $\lambda_{4,3}$; ($\triangledown$): $\lambda_{3,2}$; ($\circ$): $\lambda_{2,3}$. 
        \label{fig:fixed_points_TCR}
    }
  \end{center}
\end{figure}
The locations and eigenvalues of the fixed points of the TCR flow are calculated in appendix \ref{app:fixed_points_slip}. 
A sketch showing those stagnation points in the case of global chaos $U_1=0.25$ is given in figure \ref{fig:fixed_points_TCR}$a$: 
there are 18 stagnation points, all located on the boundary of the cavity, each with one or two directions of stability (possibly of the spiral kind, \textit{i.e.} associated with complex conjugate eigenvalues). 
Note that those fixed points exist for all cases studied here ($U_1\le0.25$), although their location may change for points of type $(3)$ and $(4)$. 
For most of the cases studied here ($U_1\le0.25$), as seen in figure \ref{fig:fixed_points_TCR}$b$, the negative eigenvalues nearer to zero are such that
\begin{equation}
0>\lambda_{3,1}\approx\lambda_{2,3}>\lambda_{4,3}>\lambda_{4,1}\, .
\label{eq:ordered_lambda}
\end{equation}

\begin{figure}
  \begin{center}
\includegraphics[clip=]{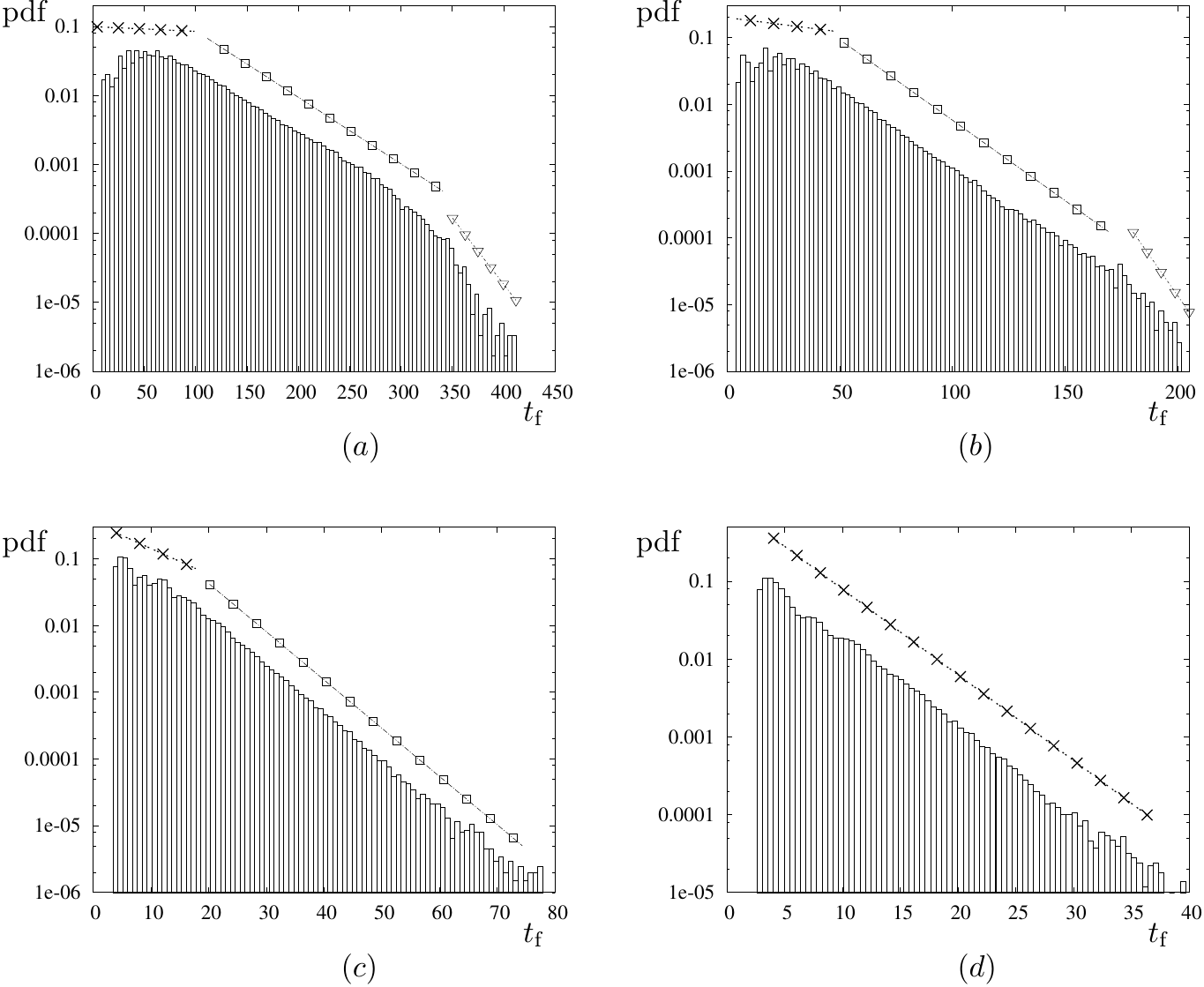}
    \caption
    {
	Distribution of time of flight for: $(a)$ $U_1=0.02$; 
	$(b)$ $U_1=0.05$; $(c)$ $U_1=0.15$; $(d)$ $U_1=0.25$.
	The symbols correspond to fits with negative eigenvalues; ($\times$): $\lambda_{3,1}$; ($\square$): $\lambda_{4,3}$; ($\triangledown$): $\lambda_{4,1}$.
        \label{fig:ToF_pdf_transadiabatic_global}
    }
  \end{center}
\end{figure}
In the case of global chaos $U_1=0.25$ (without transadiabatic drift nor elliptic fixed points), if all hyperbolic points give rise to an exponential decay, then at long times, only the decay with the negative eigenvalue nearer to zero (the one with the slower decay) should be visible. 
This is exactly what is observed in figure \ref{fig:ToF_pdf_transadiabatic_global}$d$, where the decay scales with $\lambda_{3,1}$ (equation (\ref{eq:ordered_lambda})).

When transadiabatic drift is present (here for $U_1\le0.15$), the trajectory of a given particle is almost closed (because the flow is almost regular), so that two successive intersection points $\mathbf{x_{n-1}}$ and $\mathbf{x}_n$ in the Poincar\'e section, linked by equation (\ref{eq:xn}), are very near to each other. 
This means that a fluid particle remains for a long time in a given region of the flow (where it ``visits'' some given fixed points), before visiting another region (associated with other fixed points). 
After very long times it has visited the whole domain, and it is necessary to make statistics over a very large number of Poincar\'e intersection points (here about $10^6$) for a reasonable convergence. 
As seen in figures \ref{fig:ToF_pdf_transadiabatic_global}.$a-c$, in that case the statistics are rather different than what is observed for $U_1=0.25$ (figure \ref{fig:ToF_pdf_transadiabatic_global}.$d$): 
the decay is still exponential, but not governed by a single eigenvalue (different slopes are visible in the log-lin plots). 
This particular behavior is all the more pronounced as $U_1$ is small (and the transadiabatic drift phenomenon is important). 
Moreover, the long time decay does not scale with the smallest negative exponent $\lambda_{3,1}$, but rather with $\lambda_{4,3}$, even with $\lambda_{4,1}$ for very small values of $U_1$. 
This could be explained by the fact that the particle spends longer time in regions visiting the spiraling points of type $(4)$ (figure (\ref{fig:fixed_points_TCR}.$a$)) than in the rest of the domain.
Finally, note that the \textit{shortest} times of flight seem to be rather governed by the smallest eigenvalue $\lambda_{3,1}$ for all histograms corresponding to $U_1\le 0.2$ (also on those not shown here).

\section{Summary and conclusion}
In this article we have studied different 3D chaotic stationary mixers with global chaos (no apparent regular regions nor KAM tori), and characterized them in terms of Poincar\'e sections and Lyapunov exponents. 
In the case of real mixers, the Lyapunov exponent fails in detecting the presence of solid walls, while the Poincar\'e sections do not allow to decide between walls or lines of zero normal-velocity. 
We have proposed to use the histograms of time of flight (lapse of time between two crossings of consecutive Poincar\'e sections) to study 3D chaotic systems;
this tool costs basically nothing more than the calculation of the Poincar\'e section of the flow. 
The time of flight results from a \textit{single} fluid particle that wanders in the whole chaotic region. 
However, the tail of the distribution (long times of flight) results from regions where the fluid particle remains locally trapped for a while, like in the vicinity of fixed walls, or in regions of poor stretching. 

In our numerical investigations, the histograms of time of flight reveal two very different behaviors, depending on whether the chaotic mixer possesses walls or not: 
whenever fixed solid boundaries are present, a very large tail with a $t^{-3}$ power law decay is observed, while we obtain an exponential law decay in the model case with slipping boundaries. 
We have proposed a simple model which relates this power law behavior to the presence of walls in the first case; in the case of slipping boundaries, the model shows that the exponential decay is governed by the negative eigenvalues of fixed points of the flow nearer to zero, as shown also by the numerical simulations.  

Note finally that, because mixing is also limited in the end by regions where stirring is poor, the shape of time of flight distributions could be somehow related to mixing efficiency, with an algebraic tail when scalar variance decays algebraically, and an exponential tail when scalar variance decays exponentially. 


\appendix
\acknowledgments

 \noindent\parbox[m]{0.78\textwidth}{
\quad This article would not exist without the insights and constructive analysis of my co-author and colleague Philippe, who sadly passed away on the very day on which I received its final notification of acceptance.} \hfill\parbox[m]{0.15\textwidth}{\includegraphics[clip=,width=0.15\textwidth]{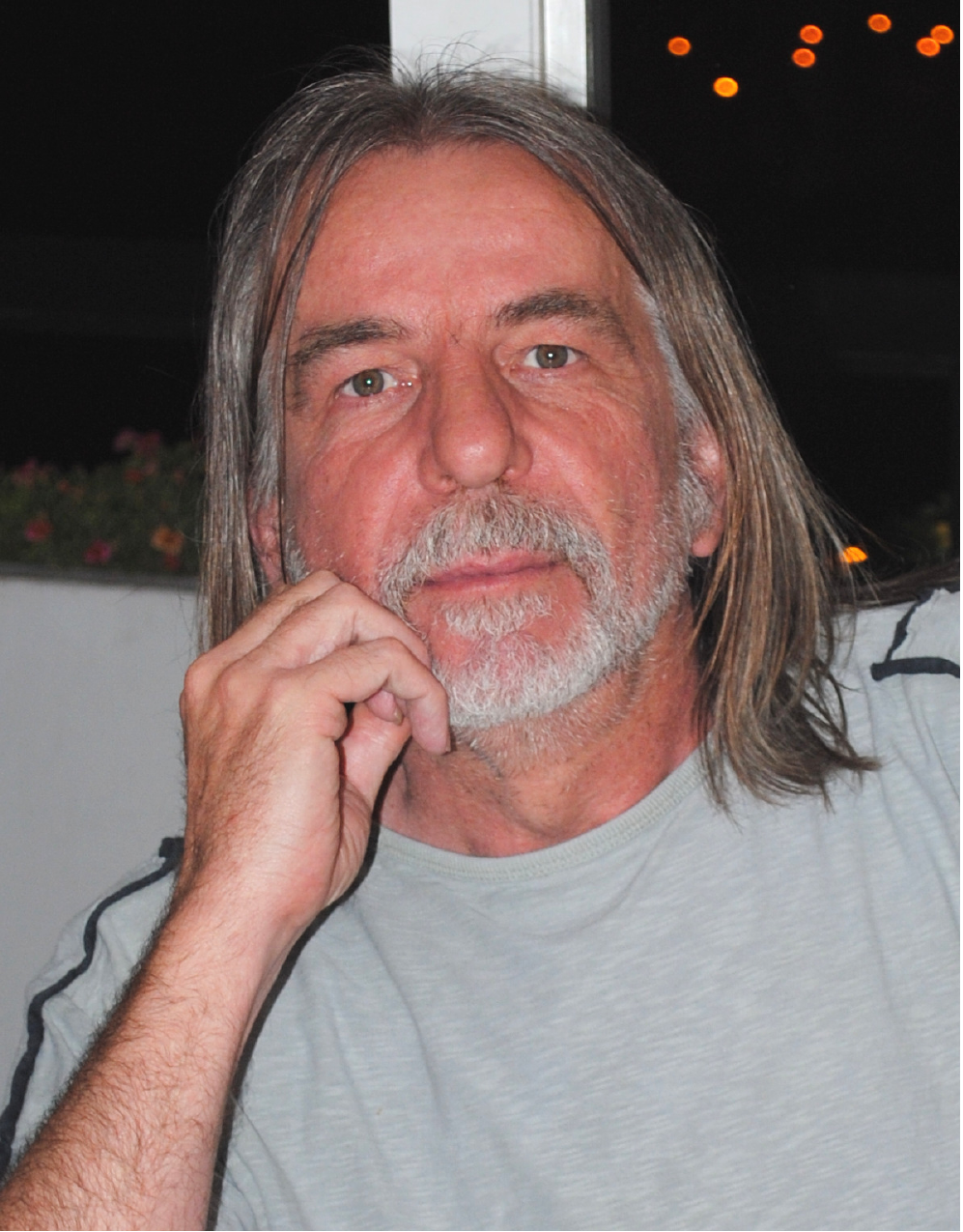}}


\section{Numerical details for the Kenics mixer}
\label{app:num_kenics}

Because Lagrangian tracking requires great accuracy, we used 368,951 pressure nodes and 2,770,011 velocity nodes (Eulerian quantities would be satisfactorily obtained with a much lower resolution).

\subsection{Inlet and outlet boundary conditions}
An important issue in open flows is the imposed boundary conditions at the inlet and the outlet.
Firstly, instead of the imposed pressure drop between inlet and outlet, we use a zero pressure drop and add a prescribed volume forcing term to momentum equations in a small part of the  domain near the inlet.
This  produces the same flow as an imposed pressure drop (Ref. \cite{bib:carriere07}).
Secondly, rather than using periodic conditions on the velocity field, we  imposed zero azimuthal and radial components and a Neumann condition on the axial component at both inlet and outlet.
Then we  checked that, owing to  the short characteristic length for establishing a Stokes flow, the values obtained for the axial components of the velocity at the outlet only slightly deviate from the ones at inlet: 
this is true to a relative error less than $0.5$~\textperthousand, which is small enough to avoid negative effects on long time integration of trajectories.

\subsection{Pressure drop}
The pressure drop, or  more properly speaking, the hydraulic resistance is an unavoidable point of comparison.
Kumar \textit{et al.} \cite{bib:kumaretal08} give some review of experimental and numerical correlations with the
Reynolds number from the literature.
Following the usual trend, we compute the ratio $K$ between the hydraulic resistance of the mixer and that of a circular pipe with equal diameter, flow rate and length. 
Even for vanishing Reynolds numbers, there is a large scattering in the results, typically $4.86$ in Ref. \cite {bib:grace71} to $7$ in Ref. \cite{bib:pahlmuschelknautz80}, while Byrde and Sawley obtained $3.59$.
Here we obtained $K = 4.67$: 
given that the depth of the blade is 2~\% of the pipe diameter (rather 5--10~\% in the industrial configuration and $0$ for Byrde and Sawley simulations), this is in accordance with the discussion in Ref. \cite{bib:byrde97} on the importance of the blade depth on the hydraulic resistance. 

\subsection{Particle tracking}
\begin{figure}
\includegraphics[clip=]{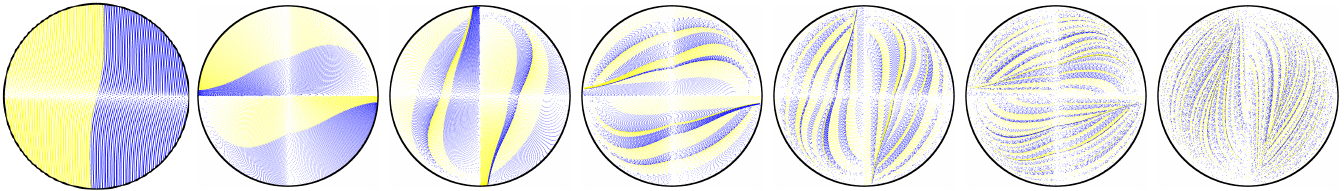}
    \caption{Fluid particles distribution in cross-planes located at, from left to right, the leading edge of the
		first elements and the end of each of the six elements for the Kenics mixer (figure 
		\ref{fig:kenics_mixer_gmtry}).
		\label{fig:Kenics_Mixer_368951_r3-2_prt}
	    }
\end{figure}

Figure \ref{fig:Kenics_Mixer_368951_r3-2_prt} shows colored particle tracking, \textit{i.e.} the distribution of marked particles in successive planes located at the end of each six elements (together with the leading edge of the first element). 
The figure may be compared favorably to the ones presented in Ref. \cite{bib:byrde97} for a Reynolds number value of
$0.01$ and also to experimental visualizations by Grace \cite{bib:grace71} (also reported in Ref. \cite{bib:byrde97} and \cite{bib:middleman77}).
The striation process appears well reproduced even if the comparison is essentially qualitative, looking like a
$2^n$ process (where $n$ is the number of mixing elements) as reported in the literature \cite{bib:grace71,bib:middleman77,bib:hobbsmuzzio97}.

\subsection{Loss of particles}
As previously mentioned, numerical limitations are especially severe in the present case. 
In a few words (see Ref. \cite{bib:carriere06} for more details), usual formulations of the discrete pressure--velocity 
problem (the so-called $P_1$-$P_2$ element \cite{bib:giraultraviart86} in the present FEM method) in three dimensions 
require some smoothness properties for the pressure field ($p \in H^1(\Omega)$ so that its derivatives must be piecewise square integrable) which cannot be satisfied in the vicinity of a ``corner", for instance.
Here, this is obviously the case near the leading edge of a blade, where visualization of the pressure field (not shown here) shows ripples of small amplitude; 
this is also the case along the entire surface of a blade (a succession of triangles), although this appears less critical.
This impacts the satisfaction of incompressibility, and explains why following a trajectory for a sufficiently long time is indeed difficult.

Hobbs and Muzzio reported about 5\% loss at the end of a 6-elements geometry, and Byrde and Sawley \cite{bib:byrdesawley99} about 1-5~\% for the tracking of 20,000 and 262,656 particles, respectively.
Here we computed the trajectory of 31,630 particles, and obtained about 0.54~\% loss (note however that it depends much on the choice of the initial location of particles).


\section{Time of flight distribution in a plane Couette flow}
\label{app:analytic_timeofflight}

\begin{figure}
  \begin{center}
\includegraphics[clip=]{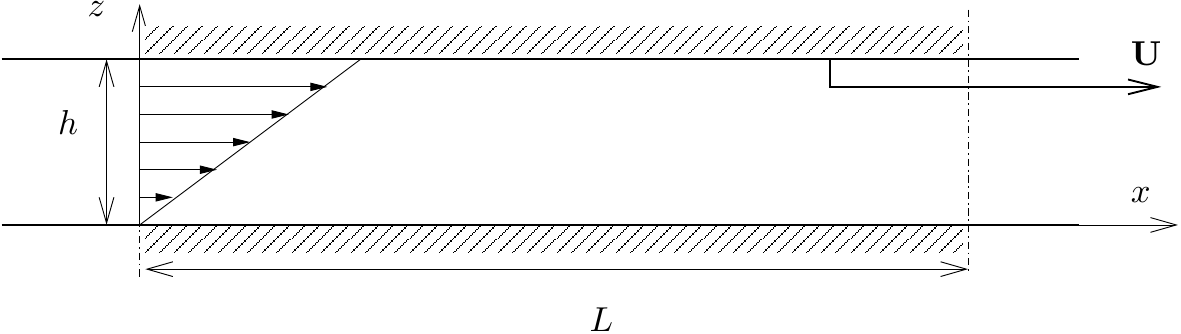}
    \caption
    {
    Plane Couette flow: the time of flight is calculated with Poincar\'e sections being separated by a length $L$.
        \label{fig:Couette}
    }
  \end{center}
\end{figure}

Let us consider the laminar flow between two parallel planes, the upper one moving at constant speed ${\bf U}=U\,{\bf e_x}$. 
The velocity-field is such that (figure \ref{fig:Couette}):
\begin{equation}
{\bf v}=v_x(z)\, {\bf e_x},\hbox{ with } v_x(z)=Uz/h,\ \hbox{with }0\le z\le h
\label{vel_Couette}
\end{equation}
where $h$ is the distance between the walls. 
As before, we let $G(t)$ the probability density to have a time of flight of duration $t$ for a section of length $L$; 
$G(t) dt$ is therefore the probability to have a time of flight in between 
$t$ and $t+dt$, which, since ${\bf v}$ depends only on $z$, is equal 
to the probability of being in between $z$ and $z+dz$, with $z$ the height
corresponding to time of flight $t$ such that 
\begin{equation}
t=L/v_x(z). 
\label{timeflight_Couette}
\end{equation}
Since the reinjection rate is proportional to the velocity, we have
\begin{equation}
G(t) dt\propto v_x(z) dz
\end{equation}
from equations  (\ref{timeflight_Couette}) and (\ref{vel_Couette}), we finally obtain:
\begin{equation}
G(t)\propto{t^{-3}}.
\end{equation}


\section{Fixed points for the slipping wall cavity flow}
\label{app:fixed_points_slip}

\subsection{location of the fixed points}
The locations of the fixed points of the slipping wall cavity flow are given by:
\begin{eqnarray}
\frac{dx}{dt}=& -U_1 \sin\pi x \cos\pi z &=0\label{dxdt}\\
\frac{dy}{dt}=& -2U_2 \sin\pi y \cos2\pi z &=0\label{dydt}\\
\frac{dz}{dt}=& U_1 \cos\pi x \sin\pi z+ U_2 \cos\pi y \sin2\pi z &=0\label{dzdt}
\end{eqnarray}
with
\begin{equation}
U_1^2+\frac{25}{4} U_2^2=1.
\end{equation}
Equations (\ref{dxdt}) and (\ref{dydt}) lead to
\begin{eqnarray}
&x=0 \hbox{ or } &x=1 \hbox{ or } z=1/2\\
&y=0 \hbox{ or } &y=1 \hbox{ or } z=1/4 \hbox{ or } z=3/4.
\end{eqnarray} 
hence: 
\begin{enumerate}
\item 8 fixed points located each at a corner of the cube;
\item 2 fixed points located at the center of two opposite walls ($x=1/2$, $y=0$, $z=1/2$ and $x=1/2$, $y=1$, $z=1/2$);
\item If $\xi_0= U_1/(2U_2)< 1$, which is equivalent to $U_1< 4/\sqrt{41}\approx0.625$, there are 4 additional points located on the vertical sides, at $x=0$, $y=0$, $z=1-z_0$ and $x=0$, $y=1$, $z=z_0$ and $x=1$, $y=1$, $z=1-z_0$ and $x=1$, $y=0$, $z=z_0$, with $\cos\pi z_0=\xi_0$;
\item If $\xi_1= U_1/(\sqrt{2}U_2)<1$, which is equivalent to $U_1<4/\sqrt{66}\approx 0.492$, there are four additional points on two opposite sides of the cubes, at $x=0$, $y=1-y_1$, $z=1/4$, and $x=0$, $y=y_1$, $z=3/4$ and $x=1$, $y=1-y_1$, $z=1/4$ and $x=1$, $y=y_1$, $z=3/4$, with $\cos\pi y_1=\xi_1$.
\end{enumerate}
Fixed points of type $i$ ($i=1,2,3$ or $4$) are indicated by $(i)$ on figure \ref{fig:fixed_points_TCR}. 
Their eigenvalues are denoted thereafter by $\lambda_{i,j}$

\subsection{Eigenvalues of the fixed points}
\subsubsection{Fixed points located at the corners ($x=\ell$, $y=m$, $z=n$, with $\ell$, $m$, and $n$ equal to $0$ or $1$)}
We let $x=\ell+X$, $y=m+Y$ and $z=n+Z$. We obtain the following linearized system for small $X$, $Y$ and $Z$:
\begin{eqnarray}
\frac{dX}{dt} &=& (-1)^{1+\ell+n}\ U_1 \pi X\\
\frac{dY}{dt} &=& (-1)^{1+m}\ 2U_2 \pi Y\\
\frac{dZ}{dt} &=& (-1)^{\ell+n}\ U_1 \pi Z+(-1)^m 2U_2 \pi Z
\end{eqnarray}
and therefore, 3 real eigenvalues:
\begin{eqnarray}
\lambda_{1,1}&=&(-1)^{1+\ell+n}\ \pi U_1\label{lambda11}\\
\lambda_{1,2}&=& (-1)^{1+m}\ 2 \pi U_2\label{lambda12}\\
\lambda_{1,3}&=& (-1)^{\ell+n}\ \pi U_1  +(-1)^m 2\pi U_2 \label{lambda13}
\end{eqnarray}

\subsubsection{Fixed points located at the center of two opposite walls ($x=1/2$, $y=m$, $z=1/2$, with $m=0$ or $1$)}
We let $x=1/2+X$, $y=m+Y$ and $z=1/2+Z$. We obtain the following linearized system for small $X$, $Y$ and $Z$:
\begin{eqnarray}
\frac{dX}{dt} &=&  U_1 \pi Z\\
\frac{dY}{dt} &=& (-1)^m\ 2U_2 \pi Y\\
\frac{dZ}{dt} &=& -U_1 \pi X+(-1)^{m+1}\ 2U_2 \pi Z
\end{eqnarray}
and therefore, 3 eigenvalues (2 of which whether real or complex depending on the sign of $U_2^2-U_1^2$):
\begin{eqnarray}
\lambda_{2,1}&=&(-1)^m\ \pi 2U_2\label{lambda21}\\
\lambda_{2,2}&=& (-1)^{1+m}\ \pi \left[U_2+(U_2^2-U_1^2)^{\frac{1}{2}}\right]\label{lambda22}\\
\lambda_{2,3}&=& (-1)^{1+m}\ \pi \left[U_2-(U_2^2-U_1^2)^{\frac{1}{2}}\right]\label{lambda23}
\end{eqnarray}

\subsubsection{Fixed points located on the vertical sides of the cube ($x=\ell$, $y=m$, $z=Z_0$, with $\ell$ or $m=0$ or $1$, and $\cos\pi Z_0=(-1)^{1+\ell+m}U_1/(2U_2)$), which exist when $U_1<4/\sqrt{41}$}
We let $x=\ell+X$, $y=m+Y$ and $z=Z_0+Z$. We obtain the following linearized system for small $X$, $Y$ and $Z$:
\begin{eqnarray}
\frac{dX}{dt} &=& (-1)^m\ \left[U_1^2/(2U_2)\right]\pi X\\
\frac{dY}{dt} &=& (-1)^m\ 2U_2 \left[1-U_1^2/(2U_2^2)\right]\pi Y\\
\frac{dZ}{dt} &=& (-1)^{m+1}\ 2U_2 \left[1-U_1^2/(4U_2^2)\right]\pi Z
\end{eqnarray}
and therefore, 3 real eigenvalues:
\begin{eqnarray}
\lambda_{3,1}&=&(-1)^m\ \pi U_1^2/(2U_2)\label{lambda31}\\
\lambda_{3,2}&=& (-1)^m\ \pi 2U_2\left[1-U_1^2/(2U_2^2)\right]\label{lambda32}\\
\lambda_{3,3}&=& (-1)^{1+m}\ \pi 2U_2\left[1-U_1^2/(4U_2^2)\right]\label{lambda33}
\end{eqnarray}

\subsubsection{Fixed points located on two opposite sides of the cube, which exist when $U_1<4/\sqrt{66}$}
We let $x=\ell+X$, $y=Y_1+Y$, $z=Z_1+Z$, with $\ell$ and $n=0$ or $1$, $Z_1$ such that $\cos\pi Z_1=(-1)^n\ \sqrt{2}/2$ and $\cos\pi Y_1=(-1)^{1+\ell+n}\ \sqrt{2}U_1/(2U_2)$. 
We obtain the following linearized system for small $X$, $Y$ and $Z$:
\begin{eqnarray}
\frac{dX}{dt} &=& (-1)^{1+\ell+n}\ \sqrt{2}/2\, U_1\ \pi X\\
\frac{dY}{dt} &=& (-1)^n\ 4U_2\ \sin\pi Y_1\ \pi Z\\
\frac{dZ}{dt} &=& (-1)^{n+1}\  U_2\ \sin\pi Y_1\ \pi Y+(-1)^{\ell+n}\ U_1\sqrt{2}/2 \pi Z
\end{eqnarray}
and therefore, 3 eigenvalues, two of which whether real or complex depending on the sign of $17/8 U_1^2-4U_2^2$:
\begin{eqnarray}
\lambda_{4,1}&=&(-1)^{1+\ell+n}\ \pi \sqrt{2}/2\, U_1\label{lambda41}\\
\lambda_{4,2}&=& (-1)^{\ell+n}\ \pi \left[U_1\sqrt{2}/4+(17/8 U_1^2-4U_2^2)^{\frac{1}{2}}\right]\label{lambda42}\\
\lambda_{4,3}&=& (-1)^{\ell+n}\ \pi \left[U_1\sqrt{2}/4-(17/8 U_1^2-4U_2^2)^{\frac{1}{2}}\right]\label{lambda43}
\end{eqnarray}


\end{document}